\documentclass[showpacs,preprintnumbers,floatfix]{revtex4}
\usepackage{graphicx}
\usepackage{dcolumn}
\usepackage{bm}
\usepackage{longtable}

\begin{document}

\title{Tables of mesoscopic 3D Coulomb balls}

\author{P.~Ludwig$^{1,2}$, S.~Kosse$^{1}$, V.~Golubnychiy$^{1}$, M.~Bonitz$^{1}$, and H.~Fehske$^{3}$}

\affiliation{$^{1}$Christian-Albrechts-Universit{\"a}t zu Kiel,
Institut f{\"u}r Theoretische Physik und Astrophysik, Leibnizstr. 15, 24098 Kiel, Germany}
\affiliation{$^{2}$Universit{\"a}t Rostock, Fachbereich Physik, l8051 Rostock, Germany}
\affiliation{$^{3}$Universit{\"a}t Greifswald, Institut f{\"u}r Physik, l7487 Greifswald, Germany}
\date{\today}

\begin{abstract}

Detailed numerical results for the structural properties of three-dimensional classical
Coulomb clusters confined in a spherical parabolic trap are presented. Based on extensive high accuracy
computer simulations the shell configurations and energies  for particle numbers in
the range $2 \le N\le 160$ are reported. Further, the  mean shell radii and shell widths
are calculated.

\end{abstract}
\pacs{52.27Gr,82.70.Dd}

\maketitle

{\em Model:}
We consider $N$ classical particles with equal charge $q$ and mass $m$ interacting via the
Coulomb force and being confined in a 3D isotropic harmonic trap with frequency $\omega$
with the hamiltonian
\begin{eqnarray}\label{hamilton}
H_N =  \sum\limits_{i=1}^{N} \frac{m}{2}\dot{r_i}^2 +
         \sum\limits_{i=1}^{N} \frac{m}{2}\omega^2 r_i^2 +
	 \sum\limits_{i>j}^{N} \frac{q^2}{4\pi\varepsilon|{\bf r}_i-{\bf r}_j|}.
\end{eqnarray}
Below we will use dimensionless lengths and energies by introducing the units $r_0 = (q^2/2 \pi \varepsilon m \omega^2)^{1/3}$
and $E_0 = (m \omega^2 q^4/32 \pi^2 \varepsilon^2)^{1/3}$, respectively.

There have been a number of numerical investigations of the system (\ref{hamilton}). Here we list
the references we are aware of and extend/correct their data up to $N=160$. A subset of this data
together with an analysis of metastable states and symmetry properties is discussed in Ref.
\cite{ludwig04}.

Rafac et al. \cite{rafac91}, correcting earlier results, identified the first shell closure at $N=12$
(the 13th particle is the first to occupy a second shell) and presented detailed data, including ground state energies for $N \le 27$, but they missed the onset of the third shell, as did Hasse et al. \cite{hasse91}. Tsuruta et al. extended the table to $N=59$ \cite{tsuruta93}. The most extensive data, for up to a few thousand particles, has been presented by Hasse et al. \cite{hasse91}, but they report excited states rather than the ground states for $N=28-31, 44, 54$ and practically for all $N>63$ (except for $N=66$). The obvious reason for the computational difficulties is the existence of a large number of excited (metastable) states which are energetically close to the ground state; with increasing $N$ this number grows exponentially whereas the energy difference rapidly vanishes. This means that
there exists no straightforward tool to identify the ground state from the large number of (local)
minima of the potential energy surface.

We also mention related work on three-dimensional clusters of particles confined to the surface of a sphere \cite{altshuler97,erber95}.

The tables below contain simulation data (details of the molecular dynamics procedure have been
given in Ref. \cite{ludwig04}) for the shell configuration, total energy per particle, mean shell radius
(starting from the outer shell) and standard deviation of radius $r_i$ (denoting the mean width of shell i,
not the error of $r_i$).

\begin{table*}
\caption{\label{tab:groundstates2-60}Shell
configurations, energy per particle for the lowest lying states,
 mean shell radii $r_{1,2,3,4}$ and widths $\sigma_{1,2,3}$ }

\begin{ruledtabular}
%\begin{center}
\begin{tabular}[t] {|c|c|c|c|c|c|c|c|c|c|}
$N$  &  Config.          &  $E/N$       &     $r_{1}$     & $r_{2}$        &  $r_{3}$        &  $r_{4}$       &  $\sigma_{1}$  &  $\sigma_{2}$   &  $\sigma_{3}$         \\ [0.3ex]  \hline
2   & (2)              & 0.500000(0)   & 0.500(0)  & -             & -              & -            & 0.000(0)       & -             & -              \\ [0.2ex]  \hline
3   & (3)              & 1.310370(7)   & 0.660(9)  & -             & -              & -            & 0.000(0)       & -             & -              \\ [0.2ex]  \hline
4   & (4)              & 1.785826(2)   & 0.771(5)  & -             & -              & -            & 0.000(0)       & -             & -              \\ [0.2ex]  \hline
5   & (5)              & 2.245187(2)   & 0.865(1)  & -             & -              & -            & 0.010(0)       & -             & -              \\ [0.2ex]  \hline
6   & (6)              & 2.654039(0)   & 0.940(6)  & -             & -              & -            & 0.000(0)       & -             & -              \\ [0.2ex]  \hline
7   & (7)              & 3.064186(0)   & 1.010(6)  & -             & -              & -            & 0.013(5)       & -             & -              \\ [0.2ex]  \hline
8   & (8)              & 3.443409(4)   & 1.071(4)  & -             & -              & -            & 0.000(0)       & -             & -              \\ [0.2ex]  \hline
9   & (9)              & 3.809782(0)   & 1.126(9)  & -             & -              & -            & 0.006(3)       & -             & -              \\ [0.2ex]  \hline
10  & (10)             & 4.164990(0)   & 1.178(3)  & -             & -              & -            & 0.005(5)       & -             & -              \\ [0.2ex]  \hline
11  & (11)             & 4.513275(4)   & 1.226(5)  & -             & -              & -            & 0.010(1)       & -             & -              \\ [0.2ex]  \hline
12  & (12)             & 4.838966(4)   & 1.270(0)  & -             & -              & -            & 0.000(0)       & -             & -              \\ [0.2ex]  \hline
13  & (12, 1)          & 5.166798(3)   & 1.365(9)  & 0.000(0)      & -              & -            & 0.000(0)       & 0.000(0)      & -              \\ [0.2ex]  \hline
14  & (13, 1)          & 5.485915(4)   & 1.403(3)  & 0.007(1)      & -              & -            & 0.005(8)       & 0.000(0)      & -              \\ [0.2ex]  \hline
15  & (14, 1)          & 5.792094(2)   & 1.438(3)  & 0.000(0)      & -              & -            & 0.005(6)       & 0.000(0)      & -              \\ [0.2ex]  \hline
16  & (15, 1)          & 6.093421(3)   & 1.471(9)  & 0.000(0)      & -              & -            & 0.005(2)       & 0.000(0)      & -              \\ [0.2ex]  \hline
17  & (16, 1)          & 6.388609(9)   & 1.504(2)  & 0.000(0)      & -              & -            & 0.006(2)       & 0.000(0)      & -              \\ [0.2ex]  \hline
18  & (17, 1)          & 6.678830(3)   & 1.535(3)  & 0.000(0)      & -              & -            & 0.000(6)       & 0.000(0)      & -              \\ [0.2ex]  \hline
19  & (18, 1)          & 6.964146(0)   & 1.565(4)  & 0.000(0)      & -              & -            & 0.004(0)       & 0.000(0)      & -              \\ [0.2ex]  \hline
20  & (19, 1)          & 7.247181(0)   & 1.594(6)  & 0.000(2)      & -              & -            & 0.006(9)       & 0.000(0)      & -              \\ [0.2ex]  \hline
21  & (20, 1)          & 7.522377(7)   & 1.622(6)  & 0.000(0)      & -              & -            & 0.003(4)       & 0.000(0)      & -              \\ [0.2ex]  \hline
22  & (21, 1)          & 7.795468(9)   & 1.649(9)  & 0.000(7)      & -              & -            & 0.006(3)       & 0.000(0)      & -              \\ [0.2ex]  \hline
23  & (21, 2)          & 8.063575(4)   & 1.707(7)  & 0.530(2)      & -              & -            & 0.030(2)       & 0.000(0)      & -              \\ [0.2ex]  \hline
24  & (22, 2)          & 8.326802(8)   & 1.732(6)  & 0.526(0)      & -              & -            & 0.029(4)       & 0.009(2)      & -              \\ [0.2ex]  \hline
25  & (23, 2)          & 8.588360(7)   & 1.757(0)  & 0.526(2)      & -              & -            & 0.026(3)       & 0.000(0)      & -              \\ [0.2ex]  \hline
26  & (24, 2)          & 8.844236(2)   & 1.780(5)  & 0.524(1)      & -              & -            & 0.026(3)       & 0.00(08)      & -              \\ [0.2ex]  \hline
27  & (24, 3)          & 9.097334(6)   & 1.830(5)  & 0.689(8)      & -              & -            & 0.036(9)       & 0.009(6)      & -              \\ [0.2ex]  \hline
28  & (25, 3)          & 9.348367(8)   & 1.852(5)  & 0.688(9)      & -              & -            & 0.036(4)       & 0.001(6)      & -              \\ [0.2ex]  \hline
29  & (25, 4)          & 9.595435(1)   & 1.899(2)  & 0.798(7)      & -              & -            & 0.037(4)       & 0.011(9)      & -              \\ [0.2ex]  \hline
30  & (26, 4)          & 9.838964(7)   & 1.919(7)  & 0.796(1)      & -              & -            & 0.034(7)       & 0.018(0)      & -              \\ [0.2ex]  \hline
31  & (27, 4)          & 10.079511(0)  & 1.939(9)  & 0.792(6)      & -              & -            & 0.038(3)       & 0.007(1)      & -              \\ [0.2ex]  \hline
32  & (28, 4)          & 10.318678(8)  & 1.959(6)  & 0.793(5)      & -              & -            & 0.033(8)       & 0.000(0)      & -              \\ [0.2ex]  \hline
33  & (29, 4)          & 10.556587(1)  & 1.979(1)  & 0.791(4)      & -              & -            & 0.034(6)       & 0.010(7)      & -              \\ [0.2ex]  \hline
34  & (30, 4)          & 10.790841(9)  & 1.998(0)  & 0.790(1)      & -              & -            & 0.035(8)       & 0.000(0)      & -              \\ [0.2ex]  \hline
35  & (30, 5)          & 11.022731(0)  & 2.038(1)  & 0.885(9)      & -              & -            & 0.041(0)       & 0.038(0)      & -              \\ [0.2ex]  \hline
36  & (30, 6)          & 11.251922(6)  & 2.077(5)  & 0.958(2)      & -              & -            & 0.035(3)       & 0.000(0)      & -              \\ [0.2ex]  \hline
37  & (31, 6)          & 11.478747(2)  & 2.094(7)  & 0.958(5)      & -              & -            & 0.035(8)       & 0.017(8)      & -              \\ [0.2ex]  \hline
38  & (32, 6)          & 11.702951(6)  & 2.111(9)  & 0.954(9)      & -              & -            & 0.039(4)       & 0.000(0)      & -              \\ [0.2ex]  \hline
39  & (33, 6)          & 11.928322(8)  & 2.128(9)  & 0.954(9)      & -              & -            & 0.035(2)       & 0.012(0)      & -              \\ [0.2ex]  \hline
40  & (34, 6)          & 12.150162(9)  & 2.145(3)  & 0.954(7)      & -              & -            & 0.038(0)       & 0.011(8)      & -              \\ [0.2ex]  \hline
41  & (35, 6)          & 12.370791(5)  & 2.161(8)  & 0.953(8)      & -              & -            & 0.035(4)       & 0.006(4)      & -              \\ [0.2ex]  \hline
42  & (35, 7)          & 12.589139(3)  & 2.196(1)  & 1.026(0)      & -              & -            & 0.040(6)       & 0.050(7)      & -              \\ [0.2ex]  \hline
43  & (36, 7)          & 12.805545(2)  & 2.211(9)  & 1.025(2)      & -              & -            & 0.037(4)       & 0.045(6)      & -              \\ [0.2ex]  \hline
44  & (36, 8)          & 13.020077(9)  & 2.245(4)  & 1.084(5)      & -              & -            & 0.038(0)       & 0.013(2)      & -              \\ [0.2ex]  \hline
45  & (37, 8)          & 13.232901(2)  & 2.260(3)  & 1.084(5)      & -              & -            & 0.038(0)       & 0.034(0)      & -              \\ [0.2ex]  \hline
46  & (38, 8)          & 13.444601(5)  & 2.275(1)  & 1.084(2)      & -              & -            & 0.036(4)       & 0.037(3)      & -              \\ [0.2ex]  \hline
47  & (38, 9)          & 13.654458(5)  & 2.306(6)  & 1.139(1)      & -              & -            & 0.034(0)       & 0.048(8)      & -              \\ [0.2ex]  \hline
48  & (39, 9)          & 13.862762(0)  & 2.321(0)  & 1.137(9)      & -              & -            & 0.033(0)       & 0.036(2)      & -              \\ [0.2ex]  \hline
49  & (40, 9)          & 14.069919(9)  & 2.335(1)  & 1.137(1)      & -              & -            & 0.034(1)       & 0.035(4)      & -              \\ [0.2ex]  \hline
50  & (41, 9)          & 14.275728(5)  & 2.349(0)  & 1.137(2)      & -              & -            & 0.036(8)       & 0.026(7)      & -              \\ [0.2ex]  \hline
51  & (41, 10)         & 14.480101(0)  & 2.378(8)  & 1.187(7)      & -              & -            & 0.035(2)       & 0.029(8)      & -              \\ [0.2ex]  \hline
52  & (42, 10)         & 14.683192(6)  & 2.392(2)  & 1.187(5)      & -              & -            & 0.034(0)       & 0.029(4)      & -              \\ [0.2ex]  \hline
53  & (43, 10)         & 14.885283(9)  & 2.405(5)  & 1.187(2)      & -              & -            & 0.037(5)       & 0.029(5)      & -              \\ [0.2ex]  \hline
54  & (44, 10)         & 15.085702(8)  & 2.418(6)  & 1.187(2)      & -              & -            & 0.035(2)       & 0.024(5)      & -              \\ [0.2ex]  \hline
55  & (43, 12)         & 15.284702(6)  & 2.461(8)  & 1.277(3)      & -              & -            & 0.031(8)       & 0.010(1)      & -              \\ [0.2ex]  \hline
56  & (44, 12)         & 15.482144(4)  & 2.474(3)  & 1.278(0)      & -              & -            & 0.036(9)       & 0.010(1)      & -              \\ [0.2ex]  \hline
57  & (45, 12)         & 15.679350(2)  & 2.486(9)  & 1.276(3)      & -              & -            & 0.036(3)       & 0.007(2)      & -              \\ [0.2ex]  \hline
58  & (45, 12, 1)      & 15.875406(2)  & 2.512(6)  & 1.376(5)      &  0.005(2)      & -            & 0.046(3)       & 0.004(3)      & 0.000(0)       \\ [0.2ex]  \hline
59  & (46, 12, 1)      & 16.070103(4)  & 2.524(7)  & 1.376(4)      &  0.000(0)      & -            & 0.048(0)       & 0.000(0)      & 0.000(0)       \\ [0.2ex]  \hline
60  & (48, 12)         & 16.263707(3)  & 2.523(6)  & 1.275(5)      &  -             & -            & 0.036(0)       & 0.003(6)      & -              \\ [0.2ex]
\end{tabular}

%\end{center}

\end{ruledtabular}
\end{table*}

\begin{table*}
\caption{\label{tab:groundstates61-120}
Continuation of Table \ref{tab:groundstates2-60} }
\begin{ruledtabular}
%\begin{center}
\begin {tabular}[t] {|c|c|c|c|c|c|c|c|c|c|}
$N$  &  Config.          &  $E/N$       &     $r_{1}$     & $r_{2}$        &  $r_{3}$        &  $r_{4}$       &  $\sigma_{1}$  &  $\sigma_{2}$   &  $\sigma_{3}$         \\ [0.3ex]  \hline
61  & (48, 12, 1)      & 16.455812(8)  & 2.548(8)  & 1.375(1)      &  0.004(2)      & -            & 0.045(1)       & 0.002(4)      & 0.000(0)       \\ [0.2ex]  \hline
62  & (48, 13, 1)      & 16.647519(7)  & 2.573(8)  & 1.413(4)      &  0.016(3       & -            & 0.044(3)       & 0.023(5)      & 0.000(0)       \\ [0.2ex]  \hline
63  & (48, 14, 1)      & 16.837694(0)  & 2.598(8)  & 1.447(3)      &  0.004(6       & -            & 0.039(3)       & 0.024(7)      & 0.000(0)       \\ [0.2ex]  \hline
64  & (49, 14, 1)      & 17.027288(9)  & 2.610(1)  & 1.447(8)      &  0.001(9       & -            & 0.037(3)       & 0.023(7)      & 0.000(0)       \\ [0.2ex]  \hline
65  & (50, 14, 1)      & 17.215360(8)  & 2.621(2)  & 1.447(7)      &  0.000(0)      & -            & 0.049(5)       & 0.018(8)      & 0.000(0)       \\ [0.2ex]  \hline
66  & (50, 15, 1)      & 17.402891(3)  & 2.645(3)  & 1.480(5)      &  0.005(9       & -            & 0.043(2)       & 0.026(6)      & 0.000(0)       \\ [0.2ex]  \hline
67  & (51, 15, 1)      & 17.589347(4)  & 2.656(3)  & 1.480(3)      &  0.004(6       & -            & 0.043(0)       & 0.024(3)      & 0.000(0)       \\ [0.2ex]  \hline
68  & (51, 16, 1)      & 17.774874(4)  & 2.679(7)  & 1.512(3)      &  0.003(4       & -            & 0.034(5)       & 0.031(1)      & 0.000(0)       \\ [0.2ex]  \hline
69  & (52, 16, 1)      & 17.959432(2)  & 2.690(3)  & 1.512(6)      &  0.001(0       & -            & 0.039(3)       & 0.034(3)      & 0.000(0)       \\ [0.2ex]  \hline
70  & (53, 16, 1)      & 18.143338(3)  & 2.701(0)  & 1.511(9)      &  0.002(3       & -            & 0.041(1)       & 0.031(7)      & 0.000(0)       \\ [0.2ex]  \hline
71  & (54, 16, 1)      & 18.326281(9)  & 2.711(6)  & 1.511(8)      &  0.008(3       & -            & 0.041(2)       & 0.028(0)      & 0.000(0)       \\ [0.2ex]  \hline
72  & (54, 17, 1)      & 18.508444(3)  & 2.734(2)  & 1.542(3)      &  0.005(9       & -            & 0.035(3)       & 0.020(1)      & 0.000(0)       \\ [0.2ex]  \hline
73  & (55, 17, 1)      & 18.689729(4)  & 2.744(5)  & 1.542(2)      &  0.004(7       & -            & 0.037(5)       & 0.020(4)      & 0.000(0)       \\ [0.2ex]  \hline
74  & (56, 17, 1)      & 18.870167(9)  & 2.754(6)  & 1.542(3)      &  0.008(8       & -            & 0.042(2)       & 0.017(8)      & 0.000(0)       \\ [0.2ex]  \hline
75  & (56, 18, 1)      & 19.049742(1)  & 2.776(5)  & 1.571(7)      &  0.005(5       & -            & 0.037(2)       & 0.031(8)      & 0.000(0)       \\ [0.2ex]  \hline
76  & (57, 18, 1)      & 19.228600(2)  & 2.786(5)  & 1.571(4)      &  0.000(0       & -            & 0.037(2)       & 0.025(3)      & 0.000(0)       \\ [0.2ex]  \hline
77  & (58, 18, 1)      & 19.406816(5)  & 2.796(4)  & 1.571(4)      &  0.003(3)      & -            & 0.038(5)       & 0.031(4)      & 0.000(0)       \\ [0.2ex]  \hline
78  & (59, 18, 1)      & 19.584175(2)  & 2.806(3)  & 1.571(5)      &  0.004(6)      & -            & 0.039(8)       & 0.027(1)      & 0.000(0)       \\ [0.2ex]  \hline
79  & (60, 18, 1)      & 19.760799(9)  & 2.816(1)  & 1.570(9)      &  0.005(0)      & -            & 0.040(2)       & 0.027(4)      & 0.000(0)       \\ [0.2ex]  \hline
80  & (60, 19, 1)      & 19.936689(9)  & 2.837(0)  & 1.600(2)      &  0.003(0)      & -            & 0.038(4)       & 0.038(4)      & 0.000(0)       \\ [0.2ex]  \hline
81  & (60, 20, 1)      & 20.111592(4)  & 2.857(7)  & 1.627(1)      &  0.006(4)      & -            & 0.031(1)       & 0.040(6)      & 0.000(0)       \\ [0.2ex]  \hline
82  & (61, 20, 1)      & 20.286103(1)  & 2.867(1)  & 1.627(4)      &  0.005(0)      & -            & 0.031(1)       & 0.040(6)      & 0.000(0)       \\ [0.2ex]  \hline
83  & (61, 20, 2)      & 20.459834(2)  & 2.886(6)  & 1.688(6)      &  0.544(7)      & -            & 0.039(0)       & 0.061(9)      & 0.044(5)       \\ [0.2ex]  \hline
84  & (61, 21, 2)      & 20.632758(9)  & 2.906(4)  & 1.714(0)      &  0.542(6)      & -            & 0.034(1)       & 0.069(2)      & 0.003(3)       \\ [0.2ex]  \hline
85  & (62, 21, 2)      & 20.804907(5)  & 2.915(6)  & 1.713(5)      &  0.542(2)      & -            & 0.038(6)       & 0.063(9)      & 0.021(7)       \\ [0.2ex]  \hline
86  & (63, 21, 2)      & 20.976517(8)  & 2.924(7)  & 1.713(8)      &  0.540(3)      & -            & 0.041(2)       & 0.061(4)      & 0.009(6)       \\ [0.2ex]  \hline
88  & (64, 22, 2)      & 21.317682(0)  & 2.953(2)  & 1.737(8)      &  0.538(5)      & -            & 0.033(9)       & 0.059(1)      & 0.005(7)       \\ [0.2ex]  \hline
89  & (65, 22, 2)      & 21.487369(1)  & 2.962(1)  & 1.737(8)      &  0.537(5)      & -            & 0.034(4)       & 0.057(0)      & 0.000(0)       \\ [0.2ex]  \hline
90  & (66, 22, 2)      & 21.656403(7)  & 2.970(9)  & 1.737(6)      &  0.535(9)      & -            & 0.037(4)       & 0.057(5)      & 0.000(0)       \\ [0.2ex]  \hline
91  & (66, 22, 3)      & 21.824823(2)  & 2.989(1)  & 1.791(6)      &  0.705(0)      & -            & 0.043(4)       & 0.066(7)      & 0.004(0)       \\ [0.2ex]  \hline
92  & (67, 22, 3)      & 21.992541(8)  & 2.997(9)  & 1.791(1)      &  0.705(2)      & -            & 0.044(4)       & 0.064(9)      & 0.015(2)       \\ [0.2ex]  \hline
93  & (66, 24, 3)      & 22.159489(7)  & 3.026(0)  & 1.836(1)      &  0.701(9)      & -            & 0.035(8)       & 0.078(1)      & 0.013(0)       \\ [0.2ex]  \hline
94  & (67, 24, 3)      & 22.325841(4)  & 3.034(7)  & 1.835(6)      &  0.700(1)      & -            & 0.034(9)       & 0.068(3)      & 0.019(6)       \\ [0.2ex]  \hline
95  & (67, 24, 4)      & 22.491878(2)  & 3.052(2)  & 1.884(8)      &  0.808(9)      & -            & 0.035(3)       & 0.067(7)      & 0.021(5)       \\ [0.2ex]  \hline
96  & (68, 24, 4)      & 22.657270(6)  & 3.060(6)  & 1.884(6)      &  0.808(3)      & -            & 0.040(8)       & 0.067(5)      & 0.033(0)       \\ [0.2ex]  \hline
97  & (69, 24, 4)      & 22.822032(2)  & 3.068(7)  & 1.884(9)      &  0.809(5)      & -            & 0.046(1)       & 0.067(8)      & 0.029(9)       \\ [0.2ex]  \hline
98  & (69, 25, 4)      & 22.986199(1)  & 3.086(4)  & 1.905(5)      &  0.808(1)      & -            & 0.035(7)       & 0.075(0)      & 0.028(0)       \\ [0.2ex]  \hline
99  & (70, 25, 4)      & 23.149758(0)  & 3.094(5)  & 1.905(6)      &  0.807(1)      & -            & 0.043(0)       & 0.072(2)      & 0.027(9)       \\ [0.2ex]  \hline
100 & (70, 26, 4)      & 23.312759(3)  & 3.111(7)  & 1.925(9)      &  0.805(5)      & -            & 0.041(7)       & 0.074(0)      & 0.022(6)       \\ [0.2ex]  \hline
101 & (70, 27, 4)      & 23.475164(4)  & 3.129(1)  & 1.945(0)      &  0.802(8)      & -            & 0.030(1)       & 0.073(1)      & 0.005(8)       \\ [0.2ex]  \hline
102 & (72, 26, 4)      & 23.637044(1)  & 3.128(0)  & 1.924(8)      &  0.805(2)      & -            & 0.043(3)       & 0.071(0)      & 0.018(9)       \\ [0.2ex]  \hline
103 & (72, 27, 4)      & 23.798274(3)  & 3.145(1)  & 1.944(3)      &  0.801(7)      & -            & 0.037(7)       & 0.071(2)      & 0.008(2)       \\ [0.2ex]  \hline
104 & (72, 28, 4)      & 23.959361(3)  & 3.161(7)  & 1.964(1)      &  0.802(1)      & -            & 0.034(5)       & 0.078(1)      & 0.001(9)       \\ [0.2ex]  \hline
105 & (73, 28, 4)      & 24.120222(9)  & 3.169(6)  & 1.964(1)      &  0.802(0)      & -            & 0.036(3)       & 0.076(8)      & 0.010(2)       \\ [0.2ex]  \hline
106 & (74, 28, 4)      & 24.280223(2)  & 3.177(3)  & 1.964(2)      &  0.802(3)      & -            & 0.038(7)       & 0.077(2)      & 0.009(2)       \\ [0.2ex]  \hline
107 & (75, 28, 4)      & 24.439665(7)  & 3.185(0)  & 1.964(0)      &  0.801(0)      & -            & 0.040(4)       & 0.074(4)      & 0.006(7)       \\ [0.2ex]  \hline
108 & (76, 28, 4)      & 24.598713(7)  & 3.192(7)  & 1.964(0)      &  0.800(7)      & -            & 0.041(0)       & 0.072(3)      & 0.005(3)       \\ [0.2ex]  \hline
109 & (77, 28, 4)      & 24.757151(3)  & 3.200(5)  & 1.963(8)      &  0.800(6)      & -            & 0.040(6)       & 0.070(5)      & 0.003(4)       \\ [0.2ex]  \hline
110 & (77, 28, 5)      & 24.915153(9)  & 3.216(3)  & 2.006(3)      &  0.896(1)      & -            & 0.043(5)       & 0.076(3)      & 0.041(5)       \\ [0.2ex]  \hline
111 & (77, 29, 5)      & 25.072584(2)  & 3.232(2)  & 2.024(9)      &  0.896(9)      & -            & 0.040(9)       & 0.081(7)      & 0.030(3)       \\ [0.2ex]  \hline
112 & (76, 30, 6)      & 25.229492(1)  & 3.255(4)  & 2.085(7)      &  0.967(0)      & -            & 0.035(8)       & 0.095(0)      & 0.041(7)       \\ [0.2ex]  \hline
113 & (77, 30, 6)      & 25.385842(0)  & 3.263(7)  & 2.083(1)      &  0.964(0)      & -            & 0.036(8)       & 0.073(2)      & 0.016(9)       \\ [0.2ex]  \hline
114 & (78, 30, 6)      & 25.541848(2)  & 3.271(1)  & 2.082(9)      &  0.964(0)      & -            & 0.036(6)       & 0.074(5)      & 0.009(6)       \\ [0.2ex]  \hline
115 & (77, 32, 6)      & 25.697308(2)  & 3.294(9)  & 2.116(2)      &  0.963(0)      & -            & 0.026(6)       & 0.077(4)      & 0.004(4)       \\ [0.2ex]  \hline
116 & (78, 32, 6)      & 25.852252(8)  & 3.302(2)  & 2.115(9)      &  0.963(3)      & -            & 0.021(8)       & 0.076(0)      & 0.004(5)       \\ [0.2ex]  \hline
117 & (79, 32, 6)      & 26.007089(4)  & 3.309(4)  & 2.115(8)      &  0.962(2)      & -            & 0.032(4)       & 0.075(3)      & 0.005(0)       \\ [0.2ex]  \hline
118 & (80, 32, 6)      & 26.161426(8)  & 3.316(7)  & 2.115(5)      &  0.961(3)      & -            & 0.028(6)       & 0.068(5)      & 0.007(1)       \\ [0.2ex]  \hline
119 & (81, 32, 6)      & 26.315442(5)  & 3.323(7)  & 2.115(6)      &  0.962(4)      & -            & 0.036(8)       & 0.069(8)      & 0.003(1)       \\ [0.2ex]  \hline
120 & (82, 32, 6)      & 26.468996(0)  & 3.330(8)  & 2.115(7)      &  0.962(0)      & -            & 0.037(4)       & 0.070(2)      & 0.004(0)       \\ [0.2ex]

\end{tabular}
%\end{center}
\end{ruledtabular}
\end{table*}

\begin{table*}
\caption{\label{tab:groundstates121-160}
Continuation of Table \ref{tab:groundstates61-120} }

\begin{ruledtabular}
%\begin{center}
\begin {tabular}[t] {|c|c|c|c|c|c|c|c|c|c|}
$N$  &  Config.          &  $E/N$       &     $r_{1}$     & $r_{2}$        &  $r_{3}$        &  $r_{4}$       &  $\sigma_{1}$  &  $\sigma_{2}$   &  $\sigma_{3}$         \\ [0.3ex]  \hline
121 & (83, 32, 6)      & 26.622118(4)  & 3.337(9)  & 2.115(4)      &  0.961(4)      & -            & 0.038(1)       & 0.067(6)      & 0.002(8)       \\ [0.2ex]  \hline
122 & (84, 32, 6)      & 26.774879(2)  & 3.344(9)  & 2.115(5)      &  0.962(0)      & -            & 0.039(8)       & 0.068(5)      & 0.003(7)       \\ [0.2ex]  \hline
123 & (83, 34, 6)      & 26.927194(9)  & 3.367(2)  & 2.149(3)      &  0.962(5)      & -            & 0.036(7)       & 0.085(6)      & 0.004(3)       \\ [0.2ex]  \hline
124 & (84, 34, 6)      & 27.079019(5)  & 3.374(1)  & 2.149(1)      &  0.962(7)      & -            & 0.034(4)       & 0.086(7)      & 0.009(6)       \\ [0.2ex]  \hline
125 & (84, 34, 7)      & 27.230457(6)  & 3.388(4)  & 2.185(0)      &  1.034(0)      & -            & 0.035(9)       & 0.085(2)      & 0.062(7)       \\ [0.2ex]  \hline
126 & (84, 35, 7)      & 27.381438(1)  & 3.402(7)  & 2.200(9)      &  1.034(1)      & -            & 0.036(9)       & 0.089(2)      & 0.067(6)       \\ [0.2ex]  \hline
127 & (85, 35, 7)      & 27.532034(0)  & 3.409(4)  & 2.201(4)      &  1.034(0)      & -            & 0.040(1)       & 0.091(0)      & 0.042(8)       \\ [0.2ex]  \hline
128 & (85, 35, 8)      & 27.682123(2)  & 3.423(5)  & 2.235(8)      &  1.092(2)      & -            & 0.040(7)       & 0.083(2)      & 0.036(8)       \\ [0.2ex]  \hline
129 & (85, 36, 8)      & 27.831888(6)  & 3.437(9)  & 2.250(2)      &  1.091(9)      & -            & 0.032(8)       & 0.081(5)      & 0.034(1)       \\ [0.2ex]  \hline
130 & (86, 36, 8)      & 27.981234(3)  & 3.444(5)  & 2.250(1)      &  1.091(7)      & -            & 0.035(2)       & 0.083(2)      & 0.048(8)       \\ [0.2ex]  \hline
131 & (87, 36, 8)      & 28.130244(0)  & 3.451(3)  & 2.249(8)      &  1.090(9)      & -            & 0.034(2)       & 0.078(5)      & 0.030(5)       \\ [0.2ex]  \hline
132 & (87, 37, 8)      & 28.278862(5)  & 3.465(1)  & 2.264(9)      &  1.090(5)      & -            & 0.034(4)       & 0.081(8)      & 0.014(0)       \\ [0.2ex]  \hline
133 & (88, 37, 8)      & 28.427061(5)  & 3.471(8)  & 2.264(2)      &  1.091(2)      & -            & 0.035(5)       & 0.085(7)      & 0.013(4)       \\ [0.2ex]  \hline
134 & (88, 37, 9)      & 28.574953(4)  & 3.485(5)  & 2.297(0)      &  1.144(0)      & -            & 0.031(2)       & 0.071(5)      & 0.054(1)       \\ [0.2ex]  \hline
135 & (88, 38, 9)      & 28.722421(1)  & 3.499(2)  & 2.311(0)      &  1.143(6)      & -            & 0.030(2)       & 0.073(9)      & 0.048(8)       \\ [0.2ex]  \hline
136 & (89, 38, 9)      & 28.869526(8)  & 3.505(4)  & 2.311(2)      &  1.144(0)      & -            & 0.031(5)       & 0.078(7)      & 0.043(7)       \\ [0.2ex]  \hline
137 & (90, 38, 9)      & 29.016328(0)  & 3.511(9)  & 2.311(0)      &  1.144(0)      & -            & 0.033(4)       & 0.078(9)      & 0.040(2)       \\ [0.2ex]  \hline
138 & (90, 39, 9)      & 29.162701(3)  & 3.525(4)  & 2.325(1)      &  1.143(3)      & -            & 0.029(9)       & 0.080(5)      & 0.037(0)       \\ [0.2ex]  \hline
139 & (91, 39, 9)      & 29.308773(6)  & 3.531(6)  & 2.325(1)      &  1.143(0)      & -            & 0.034(2)       & 0.085(7)      & 0.034(9)       \\ [0.2ex]  \hline
140 & (91, 40, 9)      & 29.454518(1)  & 3.544(9)  & 2.339(1)      &  1.142(9)      & -            & 0.029(5)       & 0.084(2)      & 0.042(4)       \\ [0.2ex]  \hline
141 & (92, 40, 9)      & 29.599899(6)  & 3.551(4)  & 2.338(7)      &  1.141(7)      & -            & 0.033(5)       & 0.075(1)      & 0.038(4)       \\ [0.2ex]  \hline
142 & (92, 40, 10)     & 29.744962(8)  & 3.564(4)  & 2.368(9)      &  1.193(0)      & -            & 0.034(1)       & 0.070(0)      & 0.045(4)       \\ [0.2ex]  \hline
143 & (93, 40, 10)     & 29.889733(5)  & 3.570(7)  & 2.368(9)      &  1.193(2)      & -            & 0.031(4)       & 0.071(4)      & 0.032(2)       \\ [0.2ex]  \hline
144 & (94, 40, 10)     & 30.034090(4)  & 3.576(9)  & 2.368(8)      &  1.193(1)      & -            & 0.033(1)       & 0.070(7)      & 0.055(2)       \\ [0.2ex]  \hline
145 & (94, 41, 10)     & 30.178106(2)  & 3.589(8)  & 2.382(5)      &  1.192(0)      & -            & 0.035(8)       & 0.071(2)      & 0.034(8)       \\ [0.2ex]  \hline
147 & (95, 42, 10)     & 30.465219(1)  & 3.608(7)  & 2.395(7)      &  1.192(3)      & -            & 0.029(9)       & 0.079(4)      & 0.039(4)       \\ [0.2ex]  \hline
148 & (96, 42, 10)     & 30.608238(9)  & 3.614(8)  & 2.395(5)      &  1.192(3)      & -            & 0.030(6)       & 0.078(8)      & 0.036(7)       \\ [0.2ex]  \hline
149 & (96, 43, 10)     & 30.750998(2)  & 3.627(3)  & 2.409(0)      &  1.192(6)      & -            & 0.032(3)       & 0.085(3)      & 0.037(6)       \\ [0.2ex]  \hline
150 & (96, 42, 12)     & 30.893383(1)  & 3.639(5)  & 2.454(1)      &  1.281(6)      & -            & 0.034(8)       & 0.079(5)      & 0.010(3)       \\ [0.2ex]  \hline
151 & (96, 43, 12)     & 31.035390(0)  & 3.652(4)  & 2.465(9)      &  1.281(4)      & -            & 0.027(1)       & 0.068(5)      & 0.014(6)       \\ [0.2ex]  \hline
152 & (96, 44, 12)     & 31.177075(2)  & 3.664(9)  & 2.478(3)      &  1.281(1)      & -            & 0.031(1)       & 0.067(4)      & 0.016(5)       \\ [0.2ex]  \hline
153 & (97, 44, 12)     & 31.318527(6)  & 3.670(8)  & 2.478(1)      &  1.281(1)      & -            & 0.028(2)       & 0.067(3)      & 0.012(9)       \\ [0.2ex]  \hline
154 & (98, 44, 12)     & 31.459632(1)  & 3.676(9)  & 2.477(7)      &  1.281(0)      & -            & 0.026(3)       & 0.062(5)      & 0.014(4)       \\ [0.2ex]  \hline
155 & (98, 44, 12, 1)  & 31.600488(0)  & 3.688(7)  & 2.504(2)      &  1.384(6)      & 0.002(2)     & 0.030(1)       & 0.079(9)      & 0.009(0)       \\ [0.2ex]  \hline
156 & (98, 45, 12, 1)  & 31.741100(1)  & 3.700(6)  & 2.516(9)      &  1.383(8)      & 0.012(7)     & 0.033(3)       & 0.087(1)      & 0.006(2)       \\ [0.2ex]  \hline
157 & (100, 44, 12, 1) & 31.881320(7)  & 3.700(4)  & 2.503(8)      &  1.383(9)      & 0.004(3)     & 0.034(0)       & 0.076(2)      & 0.006(5)       \\ [0.2ex]  \hline
158 & (100, 45, 12, 1) & 32.021293(6)  & 3.712(2)  & 2.516(6)      &  1.383(4)      & 0.004(3)     & 0.032(7)       & 0.085(2)      & 0.006(3)       \\ [0.2ex]  \hline
159 & (101, 45, 12, 1) & 32.161014(1)  & 3.718(0)  & 2.516(4)      &  1.383(7)      & 0.005(3)     & 0.031(0)       & 0.088(1)      & 0.005(9)       \\ [0.2ex]  \hline
160 & (102, 45, 12, 1) & 32.300404(8)  & 3.723(8)  & 2.516(1)      &  1.383(3)      & 0.007(3)     & 0.034(1)       & 0.082(2)      & 0.005(2)       \\ [0.2ex]
\end{tabular}
%\end{center}
\end{ruledtabular}
\end{table*}

\end{document}